\documentclass[10pt]{article}
\usepackage{amsmath}
\usepackage{amsfonts}
\usepackage{amssymb}
\begin{document}


%
%

\title{Energy-Momentum Density of\\ Gravitational Waves}

\author{Amir M. Abbassi\footnote{amabasi@khayam.ut.ac.ir}$\;$ and $\;$ Saeed Mirshekari\footnote{smirshekari@ut.ac.ir}\\
\small{Department of Physics, University of Tehran, North Kargar Ave, Tehran, Iran.}}
\date{(Dated: April 2008)}
\maketitle

\begin{abstract}
In this paper, we elaborate the problem of energy-momentum in general relativity by energy-momentum prescriptions theory. Our aim is to calculate energy and momentum densities for the general form of gravitational waves. In this connection, we have extended the previous works by using the prescriptions of Bergmann and Tolman. It is shown that they are finite and reasonable. In addition, using Tolman prescription, exactly, leads to same results that have been obtained by Einstein and Papapetrou prescriptions.\\

\end{abstract}


\section{Introduction}	

One of the oldest and most important problems in Einstein's theory of general relativity (GR) is the notion of energy. Einstein, himself, was first one who argued about this problem just after general relativity's formulation in 1916. He believed that energy and momentum are local quantities in GR.

Misner at el. [1] showed that the energy can be localized only in systems which have spherical symmetry. Cooperstock and Sarracino [2,3] proved that if energy is localizable for spherical systems, then it can be localized in any system. In 1990, Bondi [4] argued that a nonlocalizable form of energy is not allowed in GR. After Einstein who have been suggested the first energy-momentum prescription, many different prescriptions proposed by different persons [5-12]. Recently, besides energy-momentum prescriptions theory, it was suggested another solution for energy problem in GR that is in agreement with energy-momentum prescriptions theory about localization of energy, i.e. teleparallel gravity (for example see Ref.[13]). On the other hand, some people do not believe in localization of energy and momentum in GR. In addition, some physicists propose a new concept in this regard : quasilocalization (for example see Ref.[14]). Indeed there is no generally accepted definition for 
 energy and momentum in GR until now.

In fact, the problem arises when we want to extend special relativity to GR. In special relativity and even in classical mechanics we can show the differential form of energy-momentum conservation law by
\begin{equation}
T^{\nu}_{\mu,\nu}=0 \label{1},
\end{equation}
where $T^{\nu}_{\mu}$ is the symmetric energy-momentum tensor which refers to the local flux and density of energy and momentum related to matter and all non-gravitational fields such as electromagnetic field. But, in GR, by using the Bianchi's identities ($ G^{\nu}_{\mu; \nu} $) in field equations ($ G^{\nu}_{\mu}=8 \pi T^{\nu}_{\mu} $) we obtain [10]
\begin{equation}
T^{\nu}_{\mu; \nu}=\frac{1}{\sqrt{-g}}(\sqrt{-g}T^{\nu}_{\mu})_{,\nu}-\Gamma_{\mu\lambda}^{\nu}T_{\nu}^{\lambda}=0 \label{2},
\end{equation}
where $ \Gamma^{\sigma}_{\nu \sigma} $ are the connection coefficients. This means that in GR energy-momentum tensor can not satisfy the conservation law ($ T^{\nu}_{\mu, \nu}\neq0 $). Thus, we should look forward for an alternative quantity which its ordinary derivative is zero in each point of manifold, and therefore can be localized. In reality, the quantity that is actually conserved in the sense of Eq. (1) is some effective quantity $ _{eff} T^{\nu}_{\mu} $ which is given (in one variant) by Eq. (20.18) in Ref.[1] as
\begin{equation}
_{eff}T^{\nu}_{\mu}=T^{\nu}_{\mu}+t^{\nu}_{\mu} \label{3}
\end{equation}
so that includes the portion of matter and non-gravitational fields $ T^{\nu}_{\mu} $ , i.e. energy-momentum tensor plus portion of gravitational fields $ t^{\nu}_{\mu} $ , which is called  energy-momentum pseudo-tensor. In other variants, we obtain
\begin{equation}
_{eff}T^{\nu}_{\mu}=(-g)^{\frac{n}{2}}(T^{\nu}_{\mu}+t^{\nu}_{\mu}), \label{4}
\end{equation}
where $ g=det(g_{\mu\nu}) $  and  $ n $ is a positive integer indicating the weight, because the weight of $ \sqrt{-g} $  is $ +1 $ . For each of these $ _{eff}T^{\nu}_{\mu} $, Eq. (2) can be rewritten as
\begin{equation}
_{eff}T^{\nu}_{\mu,\nu}=0. \label{5}
\end{equation}

In fact, conserved quantity $ _{eff}T^{\nu}_{\mu} $ refers to the flux and density of energy and momentum of gravitational systems. This quantity $ _{eff}T^{\nu}_{\mu} $ can be written as the divergence of some ``super-potential'',$ H_{\mu}^{[\nu \lambda]} $   that is antisymmetric in its upper two indices [16] as
\begin{equation}
_{eff}T^{\nu}_{\mu}= H_{\mu ,\lambda}^{[\nu \lambda]}. \label{6}
\end{equation}
In addition, a new function like $ U_{\mu}^{\nu \lambda} $  can also play the role of $ H_{\mu}^{[\nu \lambda]} $ if it has the following conditions
\begin{equation}\label{7}
U_{\mu}^{\nu \lambda}=H_{\mu}^{[\nu \lambda]}+\Psi_{\mu}^{\nu \lambda} 
\qquad
(\Psi_{\mu ,\lambda}^{\nu \lambda}\equiv0	\; \textit{or} \; 	\Psi_{\mu ,\lambda\nu}^{\nu \lambda}\equiv0). 
 \end{equation}
then, the quantity  $ \Theta_{\mu}^{\nu } $ which is defined by this new super-potential remains conserved locally:
\begin{equation}
\Theta_{\mu}^{\nu }=U_{\mu ,\lambda}^{\nu \lambda}\Longrightarrow \Theta_{\mu ,\nu}^\nu =0.
\end{equation}
Using this freedom on the choice of the super-potential, authors like Einstein and Tolman arrived through different methods at the following super-potentials [16]:
\begin{equation}
H_{\mu}^{[\nu\beta]}=\dfrac{1}{2\kappa}\tilde{g}_{\mu\lambda}(\tilde{g}^{\lambda\nu}\tilde{g}^{\alpha\beta}-\tilde{g}^{\lambda\beta}\tilde{g}^{\alpha\nu})_{,\alpha}\; \;(\texttt{Einstein}),
\end{equation}
\begin{equation}
\tau_{\mu}^{\nu\beta}=H_{\mu}^{[\nu\beta]}+\dfrac{1}{2\kappa}(\delta_{\mu}^{\beta}\tilde{g}^{\alpha\nu}-\delta_{\mu}^{\alpha}\tilde{g}^{\beta\nu})_{,\alpha}\;\;(\texttt{Tolman}).
\end{equation}

As mentioned above, $ t_{\mu}^{\nu} $  in Eq. (3) is not a tensor but a pseudo-tensor, a two-index object that transforms differently than the components of a tensor. Unlike a tensor, a pseudo-tensor can vanish at a point in one coordinate system but not in others. Despite this apparent defect, pseudo-tensors can be quite useful, especially in gravitational waves research (see for example [17]). The reason is that, despite their non-covariance, the $ _{eff}T_{\mu}^{\nu} $ can be used to compute covariant conserved quantities. For example, one can compute the total 4-momentum of a system that resides alone in asymptotically flat space-time by the volume integral
\begin{equation}
P^{\mu}=\int_{eff}T_{0}^{\mu}d^{3}x,
\end{equation}
where  $ d^{3}x=dx^{1} dx^{2} dx^{3} $ is a 3-volume element of constant time. Even though the integrand depends highly on one's choice of coordinates, $ P^{\mu} $ is a true vector that resides in the asymptotically flat region [15]. So, for a long time it was believed that the results are meaningful only for energy distribution of the asymptotically Minkowskian space-times. But, the interesting point is that recent investigations of Rosen and Virbhadra [18], Virbhadra [19], Aguirregabiria et al. [20], and Xulu [21], [22] showed that many energy momentum prescriptions can give the same and appealing results even for asymptotically non-flat space-times.

Considering above discussion, there are many prescriptions for new energy-momentum density ($ _{eff}T_{\mu}^{\nu} $) which their difference is a curl term. The most important cases proposed by Einstein [8], Tolman [5], Papapetrou [6], Bergmann [7], M\o ller [8], Weinberg [10], and Landau and Lifshitz [11]. Each of them has its advantages and disadvantages; it is not proved any preferences between them. However, Palmer [23] and Virbhadra [24] discussed the importance of Einstein's energy-momentum prescription and Lessner [25] believed that M\o ller's prescription is a powerful tool for calculating the energy-momentum in GR. It should be considered that for all mentioned prescriptions, one gets physically meaningful results only in Cartesian coordinates [17-19]; but M\o ller's prescription is coordinate independent.

Cooperstock [2] argued that in GR energy and momentum are localized in regions of the non-vanishing energy-momentum tensor. So, he believed that gravitational waves do not carry energy and momentum in vacuum. Thus the existence of gravitational waves was questioned. However, the theory of GR indicates the existence of these waves as solutions of Einstein's field equations [1]. In fact this problem arises because energy is not well defined in GR. Scheidegger [26] raised doubts whether gravitational radiation has well-defined existence. Rosen [27] used the Einstein and Landau-Lifshitz's prescriptions in cylindrical polar coordinates to calculate the energy-momentum densities of cylindrical gravitational waves. His results confirmed the conjecture of Scheidegger that a physical system can not radiate gravitational energy. Two years later, Rosen [28] realized the mistake and carried out the calculations in Cartesian coordinates. He found that energy and momentum densities are non-vanishing and reasonable. Then, Virbhadra [18, 29] explicitly calculated energy and momentum densities of cylindrical gravitational waves in the Einstein, Landau-Lifshitz, Tolman, and Papapetrou's prescriptions and obtained the finite, well-defined and same results by using Cartesian coordinates in all of these prescriptions.

Recently, Sharif and Azam [30] elaborated the problem of energy-momentum in GR with the help of some well-known solutions including gravitational waves. They calculated the energy and momentum densities for general form of gravitational wave's space-time, by using Einstein, Landau-Lifshitz, Papapetrou, and M\o ller's prescriptions. They obtained finite and well-defined quantities for these densities. In this paper we extend their work by using Bergmann and Tolman's prescriptions for calculating the energy-momentum densities of general form of gravitational wave's space-time. In addition, we show that the Tolman's prescription leads to the same results in comparison with previous results obtained by Einstein and Papapetrou's prescriptions.

This paper is organized as follows: in section 2 we introduce the Einstein, Tolman and Bergmann's prescriptions that we need them in our calculations. In section 3, using the prescriptions introduced in section 2, we write our results about energy and momentum densities of general form of gravitational waves. Then, discuss about plane gravitational waves as an example. We conclude in section 4, with some commands and discussions.

\textit{Conventions} : We use geometrized units in which the speed of light in vacuum $ c $ and the Newtonian gravitational constant $ G $ are taken to be equal to $ 1 $, the metric has signature $ + - - - $ , and both Latin and Greek indices may take values 0-3 or 1-3 depending on each case that have been specified. 

\section{Energy-Momentum Prescriptions}

In this paper for obtaining energy and momentum densities we shall use Einstein, Bergmann and Tolman's prescriptions. So, in this section we briefly give their needed formulae. Interested reader can refer to the mentioned references for details.
\subsection{Einstein Prescription}
The anti-symmetric energy-momentum pseudo-tensor of Einstein [8] is 
\begin{equation}
\Theta_{i}^{k}=\dfrac{1}{16\pi}H_{i ,l}^{k l},
\end{equation}
where
\begin{equation}
H_{i}^{k l}=-H_{i}^{l k}=\dfrac{g_{in}}{\sqrt{-g}}[(-g)(g^{kn}g^{lm}-g^{ln}g^{km})]_{,m}.
\end{equation}
$\Theta_{0}^{0}$, $\Theta_{\alpha}^{0}$, and $\Theta_{0}^{\alpha}(\alpha=1,2,3)$ are the energy, momentum, and energy current density components. $ \Theta_{i}^{k} $ satisfies the local conservation laws:
\begin{equation}
\dfrac{\partial\Theta_{i}^{k}}{\partial x^{k}}=0.
\end{equation}
\subsection{Bergmann Prescription}
The non-symmetric energy-momentum prescription of Bergmann [7] is given by
\begin{equation}
B^{ik}=\dfrac{1}{16\pi}\beta^{ikm}_{,m}=0,
\end{equation}
where
\begin{equation}
\beta^{ikm}=g^{ir}\nu_{r}^{km}
\end{equation}
with
\begin{equation}
\nu_{i}^{kl}=-\nu_{i}^{lk}=\dfrac{g_{in}}{\sqrt{-g}}[(-g)(g^{kn}g^{lm}-g^{ln}g^{km})]_{,m}.
\end{equation}
$ B^{00} $ and $ B^{\alpha0} (\alpha=1,2,3) $ are the energy and energy current (momentum) density components. It is to be noted that  $ B^{ik} $ satisfies the local conservation laws:
\begin{equation}
\dfrac{\partial B_{i}^{k}}{\partial x^{k}}.
\end{equation}
\subsection{Tolman Prescription}
The non-symmetric energy-momentum prescription of Tolman [5] is defined as
\begin{equation}
\tau_{i}^{k}=\dfrac{1}{8\pi}U_{i ,l}^{kl},
\end{equation}
where
\begin{equation}
U_{i}^{kl}=\sqrt{-g}[-g^{pk}V_{ip}^{l}+\dfrac{1}{2}g_{i}^{k}g^{pm}V_{pm}^{l}]
\end{equation}
with
\begin{equation}
V_{jk}^{i}=-\Gamma_{jk}^{i}+\dfrac{1}{2}g_{j}^{i}\Gamma_{mk}^{m}+\dfrac{1}{2}g_{k}^{i}\Gamma_{mj}^{m}.
\end{equation}
$ \Gamma^{i}_{jk} $ are metric connection coefficients, and $ \tau_{0}^{0} $, $ \tau_{\alpha}^{0} $, and  $ \tau_{0}^{\alpha} $ are the energy, momentum, and energy current density components respectively. It is to be noted that $\tau_{i}^{k}$  satisfies the local conservation laws:
\begin{equation}
\dfrac{\partial\tau_{i}^{k}}{\partial x^{k}}=0.
\end{equation}

\section{Gravitational Waves}

The general form of gravitational wave's line element [31] is given by
\begin{equation}
ds^{2}=e^{-M}(dt^{2}-dx^{2})-e^{-U}(e^{-V}dy^{2}+e^{V}dz^{2}),
\end{equation}
where $ U $, $ V $ and $M$ are functions of $ t $ and $ x $ only. Two examples of this form of gravitational waves were mentioned in [30]. Plane gravitational wave is an important example in this regard. Now we calculate energy and momentum densities for line element introduced in Eq. (23).

Using Einstein prescription (Eq. (12) and Eq. (13)), after some calculations we obtain the following components of energy, momentum and energy current densities
\begin{equation}
\Theta_{0}^{0}=\frac{e^{-U}}{8\pi}({\buildrel\prime\prime\over{U}}-\buildrel\prime\over{U}^{2}),
\end{equation}
\begin{equation}
\Theta_{1}^{0}=-\Theta_{0}^{1}=\frac{e^{-U}}{8\pi}(\acute{\dot{U}}-{\buildrel\prime\over{U}}\dot{U}),
\end{equation}
\begin{equation}
\Theta_{2}^{0}=\Theta_{3}^{0}=\Theta_{0}^{2}=\Theta_{0}^{3}=0,
\end{equation}
where dot and prime denote derivative related to $ t $ and $ x $  respectively. It should be noted that Sharif and Azam [30] have reached to different quantities for $ \Theta_{0}^{0} $ and $ \Theta_{0}^{1} $ that probably arises from some mistakes in their calculations.

For Bergmann prescription (Eq. (15)-(17)) we obtain the following quantities for energy-momentum densities
\begin{equation}
B_{0}^{0}=\frac{e^{-U}}{8\pi}({\buildrel\prime\over{U}}{\buildrel\prime\over{M}}+{\buildrel\prime\prime\over{U}}-\buildrel\prime\over{U}^{2}),
\end{equation}
\begin{equation}
B_{1}^{0}=\frac{e^{-U}}{8\pi}(\dot{U}\acute{M}-\dot{U}\acute{U}+\acute{\dot{U}}),
\end{equation}
\begin{equation}
B_{0}^{1}=-\frac{e^{-U}}{8\pi}(\dot{M}\acute{U}-\dot{U}\acute{U}+\acute{\dot{U}}),
\end{equation}
\begin{equation}
B_{2}^{0}=B_{3}^{0}=B_{0}^{2}=B_{0}^{3}=0.
\end{equation}

Finally, using Tolman's prescription (Eqs. (19), (20), and (21)) lead to the following expressions for energy and momentum and current density for general form of gravitational waves (Eq. (23))
\begin{equation}
\tau_{0}^{0}=\frac{e^{-U}}{8\pi}({\buildrel\prime\prime\over{U}}-\buildrel\prime\over{U}^{2}),
\end{equation}
\begin{equation}
\tau_{1}^{0}=-\tau_{0}^{1}=\dfrac{e^{-U}}{8\pi}(\acute{\dot{U}}-\acute{U}\dot{U}),
\end{equation}
\begin{equation}
\tau_{2}^{0}=\tau_{3}^{0}=\tau_{0}^{2}=\tau_{0}^{3}=0.
\end{equation}
Until now, we have found energy-momentum distribution for a general line element of the gravitational waves. In the next section we compare these results with other prescriptions' results.

\subsection{An Example: Plane Waves}

As an example of general line element of Eq. (23) we consider plane gravitational waves with following line element [1]:
\begin{equation}
ds^{2}=dt^{2}-dx^{2}-L^{2}(t,x)[e^{2\beta(t,x)}dy^{2}+e^{-2\beta(t,x)}dz^{2}],
\end{equation}
where $ L $ and $ \beta $ are arbitrary functions which satisfy the vacuum Einstein equations
\begin{equation}
L_{,\alpha\alpha}+L\beta^{2}_{,\alpha}=0,\quad( \alpha=0,1 ).
\end{equation}
Comparing this line element (Eq. (35)) with general line element of gravitational waves (Eq. (23)), one can find the functions $ M $, $ U $, and $ V $  easily
\begin{equation}
M=0,\quad U=-2\ln L,\quad V=-2\beta.
\end{equation}
The energy-momentum distribution for the plane gravitational waves can be found by substituting the corresponding values of $ M $, $ U $, and $ V $ in related equations (Eq. 24-34 ).

\section{Summary and Discussion}

In this paper we obtained the energy, momentum, and current densities for general form of gravitational waves by Einstein, Bergmann, and Tolman's prescriptions. Last year, Sharif and Azam calculated these quantities in the same space-time by other prescriptions (Einstein, Papapetrou, M\o ller, and Landau-Lifshitz). The summary of the all of these results for energy densities obtained by different prescriptions can be given by Table I. This work is one of a series of studies by the authors on energy-momentum prescriptions in general relativity~\cite{MA2, MA3, MA4}.\\

\begin{table}
\begin{center} 
\begin{tabular}{|c|c|}

\hline 
&\\Energy-Momentum Prescription & Energy Density \\ 

\hline 
&\\Einstein &  $ \Theta_{0}^{0}=\frac{e^{-U}}{8\pi}({\buildrel\prime\prime\over{U}}-\buildrel\prime\over{U}^{2})$ \\ 

\hline &\\
Papapetrou & $ \Sigma^{00}=\frac{e^{-U}}{8\pi}({\buildrel\prime\prime\over{U}}-\buildrel\prime\over{U}^{2})$ \\ 

\hline &\\
Tolman & $ \tau_{0}^{0}=\frac{e^{-U}}{8\pi}({\buildrel\prime\prime\over{U}}-\buildrel\prime\over{U}^{2})$ \\ 

\hline &\\
M\o ller &   $M_{0}^{0}=\frac{e^{-U}}{8\pi}({\buildrel\prime\over{M}}{\buildrel\prime\over{U}}-{\buildrel\prime\prime\over{M}})$ \\ 

\hline &\\
Landau-Lifshitz &  $L^{00}=\frac{e^{-2U}}{8\pi}({\buildrel\prime\prime\over{U}}-2\buildrel\prime\over{U}^{2})$  \\  

\hline &\\
Bergmann &  $B_{0}^{0}=\frac{e^{-U}}{8\pi}({\buildrel\prime\over{U}}{\buildrel\prime\over{M}}+{\buildrel\prime\prime\over{U}}-\buildrel\prime\over{U}^{2})$  \\ 

\hline 
\end{tabular} 
\end{center}
\caption{Energy densities related to gravitational wave's general line element.}
\end{table}

 Regarding contents of table1 we can extract some interesting points from it:
\begin{itemize}
 \item It is concluded that the energy-momentum densities turn out to be finite and well-defined in all prescriptions for this space-time. This result is in contradiction with Cooperstock's  viewpoint that energy and momentum are localized only in regions of the non-vanishing energy-momentum tensor. This result can be an evidence that gravitational waves are carriers of energy and momentum in vacuum.
 \item There is no dependence on function $ V $ in energy densities.
 \item We can find that three prescriptions, i.e., Einstein, Papapetrou, and Tolman prescriptions provide the same energy densities. This justifies the viewpoint of Virbhadra and his collaborators [19] that different energy-momentum prescriptions may provide some basis to define a unique quantity. However, the remaining three prescriptions give different energy densities (because of non-covariant property of pseudo-tensors). 
 \item In a special and also common case, when $ M $ is not a function of $ x $, it is not hard to show that four prescriptions (Einstein, Papapetrou, Tolman, and Bergmann) lead to the same result for energy density. This is also a stronger confirmation for Virbhadra's view point.
 \item In the previous situation ($M=M(t)$) using M\o ller's prescription leads to a zero quantity for energy density. This result is in contradiction with Lessner's viewpoint [24]. He believed that M\o ller's prescription is a powerful tool for calculating the energy-momentum pseudo-tensors in GR.
 \item If the line element has a special symmetry, i.e. $ M(t,x)=-U(t,x) $, the four above prescriptions (Einstein, Papapetrou, Tolman, and M\o ller) give the same results.
\end{itemize}

\end{document}